\documentclass[%
 reprint,
superscriptaddress,
 amsmath,amssymb,
 aps,
]{revtex4-2}

\usepackage{graphicx}
\usepackage{CJK}
\usepackage{dcolumn}
\usepackage{bm}
\usepackage{siunitx}
\usepackage{xcolor}
\usepackage{amsmath}

\begin{document}


\title{Observation of oscillating $g$-factor anisotropy arising from strong crystal lattice anisotropy in GaAs spin-3/2 hole quantum point contacts}

\author{K. L. Hudson}  \email{k.hudson@unsw.edu.au}
 \affiliation{School of Physics, University of New South Wales}
 \affiliation{ARC Centre of Excellence for Future Low-Energy Electronics Technologies (FLEET)}

\author{A. Srinivasan}%
 \affiliation{School of Physics, University of New South Wales}
 \affiliation{ARC Centre of Excellence for Future Low-Energy Electronics Technologies (FLEET)}
 
\author{D. S. Miserev}%
 \affiliation{University of Basel, Switzerland}

\author{Q. Wang}%
 \affiliation{School of Physics, University of New South Wales}
 \affiliation{ARC Centre of Excellence for Future Low-Energy Electronics Technologies (FLEET)}

\author{O. Klochan}%
 \affiliation{School of Physics, University of New South Wales}
 \affiliation{ARC Centre of Excellence for Future Low-Energy Electronics Technologies (FLEET)}

\author{O. P. Sushkov}
 \affiliation{School of Physics, University of New South Wales}
 \affiliation{ARC Centre of Excellence for Future Low-Energy Electronics Technologies (FLEET)}

\author{I. Farrer}%
\affiliation{University of Sheffield, United Kingdom}

\author{D. A. Ritchie}%
\affiliation{Cavendish Laboratory, Cambridge, United Kingdom}

\author{A. R. Hamilton}%
 \affiliation{School of Physics, University of New South Wales}
 \affiliation{ARC Centre of Excellence for Future Low-Energy Electronics Technologies (FLEET)}
 \homepage{}

\date{\today}

\begin{abstract}

Many modern spin-based devices rely on the spin-orbit interaction, which is highly sensitive to the host semiconductor and depends on the crystal orientation, crystal asymmetry, and quantum confinement asymmetry. One-dimensional quantum point contacts are a powerful tool to probe both energy and orientation dependence of the spin-orbit interaction through the effect on the hole $g$-factor. Here we investigate the role of cubic crystal asymmetry in GaAs when the quantum point contact is rotated with respect to the crystal axes. Unexpectedly the in-plane $g$-factor is found to be extremely sensitive to point contact orientation, changing by a factor of $5$ when rotated by $45^{\circ}$. This exceptionally strong crystal lattice anisotropy of the in-plane Zeeman splitting cannot be explained within axially symmetric theoretical models. Theoretical modelling based on the combined Luttinger, Rashba and Dresselhaus Hamiltonians reveals new spin-orbit contributions to the in-plane hole $g$-factor and provides excellent agreement with experimental data.

\begin{description}
\item[Usage]
\item[PACS numbers]
\end{description}
\end{abstract}

\pacs{Valid PACS appear here}
\maketitle


Quantum confined semiconductor hole systems are attracting significant research interest due to the strong spin-orbit interaction in the valence band~\cite{Winkler03}. Strong, tunable and remarkably diverse spin-orbit interaction in 2DHG spawned a new generation of quantum devices with all-electrical control of the hole spin, 
that already found applications in spin-based transistors, spin filters, spin-flip tunnelling, and spin qubits and read-out of spin qubits via Pauli spin blockade~\cite{LiNL15, WangNL16, MaurandNatComm16, HongPRL18, HendrickxNature18}, helical spin states~\cite{SauPRL10, OregPRL10, GoulkoPRL14, HudsonNatComm21} and topological states in one-dimensional wires~\cite{DattaAPL90, LossPRA98, WolfSci01, Awschalom02}. These applications require a detailed understanding of the nature of the spin-orbit interaction, which is much more complex for spin-3/2 holes than for spin-1/2 electrons. 

To date most experimental studies have focussed on spin-orbit terms due to crystal inversion asymmetry (Dresselhaus) and structural inversion asymmetry (Rashba). Here we go further, and examine the effects of cubic crystal lattice anisotropy on the spin-orbit interaction. However disentangling different contributions to the spin-orbit interaction (SOI) in experimental data is a challenging task. Quantum point contacts (QPCs) are a powerful platform for probing the spin properties of a crystalline semiconductor systems, as the 1D $g$-factor can be directly extracted from transport measurements and related to SOI terms in the Hamiltonian. We use QPCs fabricated on (100) zinc blende GaAs/AlGaAs heterostructures to quantitatively study the crystal orientation dependence of the spin-orbit interaction through its effect on the Zeeman splitting of the 1D subbands.

In a quantum confined system without SOI spin and momentum are not coupled, therefore the $g$-factor is isotropic in a 2D quantum well, a 1D quantum wire or a 0D quantum dot. 
 In the presence of SOI, confining the particles to 2D leads to a large out-of-plane $g$-factor and a small in-plane $g$-factor due to the quantization of $k_z$~\cite{WinklerPRL00, MartinAPL08, MartinPRB10, NichelePRL14}. Further confinement to a 1D quantum wire gives rise to an anisotropy in the in-plane $g$-factors, which has been intensely studied in GaAs spin-3/2 hole QPCs~\cite{DanneauPRL06, KlochanNJP09, KoduvayurPRL08, KomijaniEPL13}. In these studies the relative size of the $g$-factor with the magnetic field applied in-plane and parallel to the current ($g_\parallel$) is up to an order of magnitude larger than when the magnetic field is applied in-plane and perpendicular to the current ($g_{\perp}$). This anisotropy, which cannot be explained with purely structural inversion asymmetry, has recently been explained using a Luttinger based 2D planar model described in~\cite{MiserevPRL17,SrinivsasanPRL17} that does not take into account the crystal lattice anisotropy.  

In this paper we go further and investigate the ratio $g_\perp/g_\parallel$ for different orientations of hole QPCs in $(100)$ GaAs/AlGaAs heterostructure. We use three QPCs fabricated along $[011]$, $[010]$ and $[001]$ crystal directions, and find extremely large variation of this ratio ranging from $g_\perp/g_\parallel \approx 0.15$ in $[011]$ QPC to $g_\perp/g_\parallel \approx 2$ in $[010]$ and $[001]$ QPCs. To understand such dramatic dependence of $g_\perp/g_\parallel$ on the QPC orientation, we perform analytical and numerical calculations based on the combined Luttinger, Dresselhaus and Rashba Hamiltonians with realistic wafer parameters. We find new strongly anisotropic spin-orbital contributions to the in-plane Zeeman splitting (going beyond Refs ~\cite{WinklerPRL00, LuPRL98}) that provide excellent agreement with our data, see Fig.{\color{blue}~4b}. 

\textit{Experimental details}
Three lithographically identical QPCs with dimensions $\SI{300}{\nano\meter}\times \SI{300}{\nano\meter}$ are fabricated along crystal directions $[001]$, $[011]$ and $[010]$ on a single chip as shown in Fig.{\color{blue}~1a}. The QPCs are defined by electron beam lithography in the form of split gates deposited on the surface of the wafer. The QPCs are fabricated on an undoped $(100)$ GaAs/AlGaAs heterostructure (wafer W917) operated in accumulation mode. A negative voltage bias on the metal top gate forms a two-dimensional hole gas (2DHG) at the GaAs/AlGaAs interface $\SI{60}{\nano\meter}$ below the surface. The 2DHG was operated at a density $p = 2\times10^{11}\SI{}{\per\square\centi\meter}$, mobility $\mu = 5.5\times10^5 \SI{}{\square\centi\meter\per\volt\per\second}$, and mean free path length $l_{mfp} \approx \SI{4.1}{\micro\meter}$ (assuming $m^* = 0.2m_e$).
    
Measurements were performed in a dilution refrigerator with a superconducting magnet at a temperature of $\SI{130}{\milli\kelvin}$. An in-situ rotation system~\cite{YeohRSI10} allowed all measurements to be performed during a single cooldown.
    
Fig.{\color{blue}~1b} shows conductance measurements as a function of split gate voltage for all three QPCs. Clear quantized conductance plateaus are observed, indicating ballistic 1D transport.

\begin{figure}
    \centering
    \includegraphics[width=0.45\textwidth]{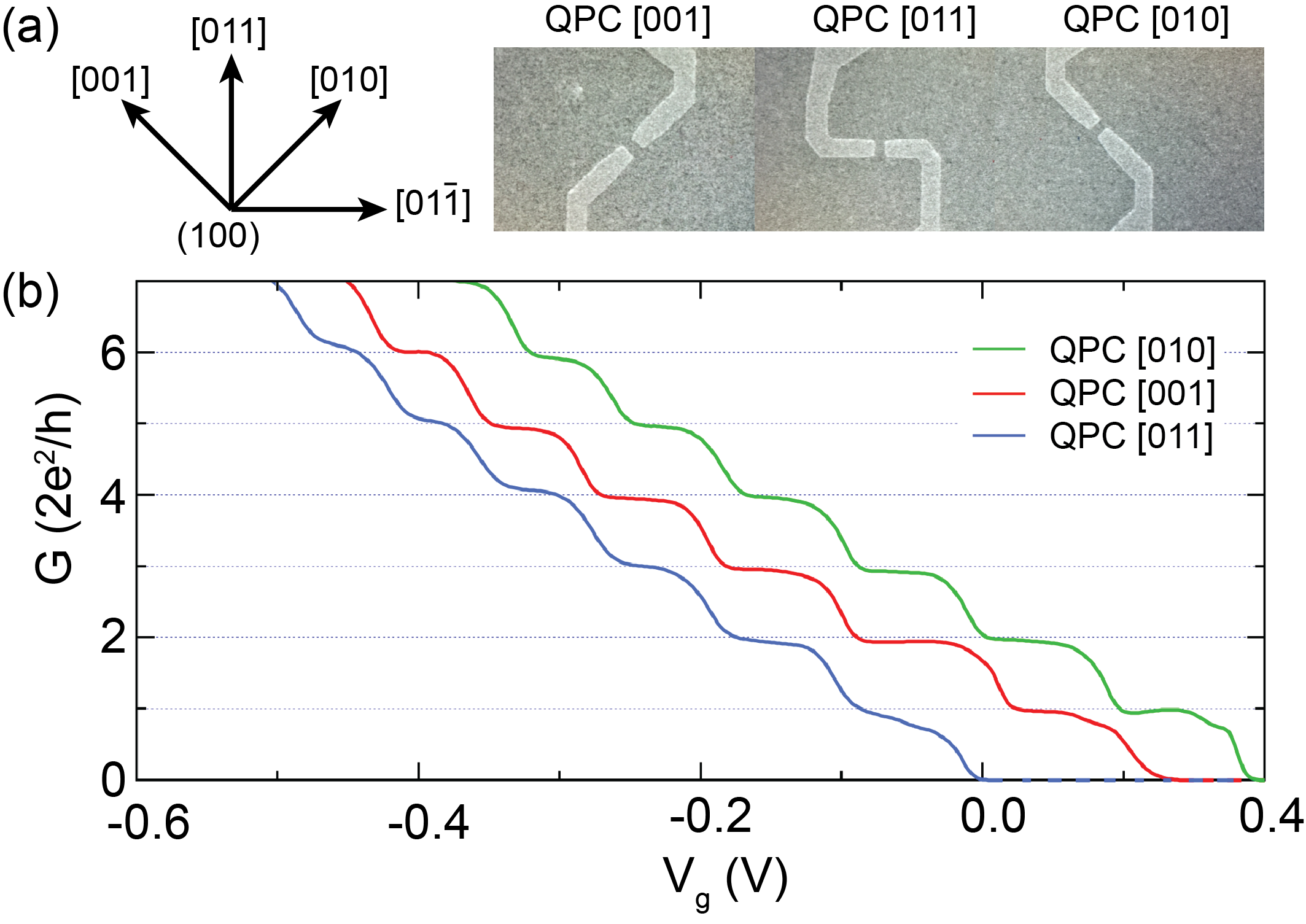}
    \caption{(a) SEM image of the device showing three quantum point contacts (QPCs) with current orientated along the [001], [011] and [010] directions on the $(100)$ crystal plane of GaAs. The lighter regions are the split gates. (b) QPC conductance as a function of split-gate voltage $V_{g}$, showing quantised plateaus in integer multiples of $2e^2/h$ with respect to gate-controlled confinement $V_g$, indicating 1D ballistic transport in each QPC.} 
    \label{fig1}
\end{figure}

\onecolumngrid

\begin{figure}[hb]
    \centering
    \includegraphics[width=\textwidth]{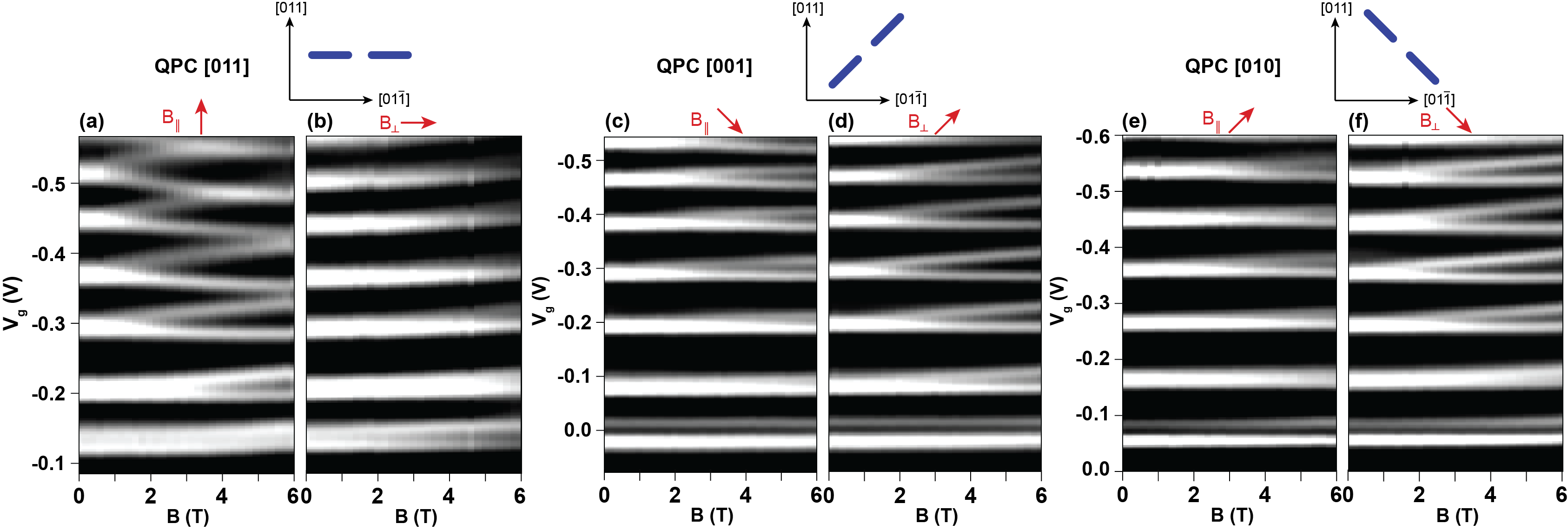}
    \caption{Greyscale maps of the transconductance $dG/dV_g$ showing Zeeman spin-splitting of the 1D hole subbands in applied magnetic field for the three QPCs (shown in Fig.{\color{blue}~1a}\color{black}). Dark regions correspond to conductance plateaus and light regions correspond to risers or subband edges. (a-b) QPC $[011]$ with magnetic field parallel $B_{\parallel}$ and perpendicular $B_{\perp}$ to the direction of current, respectively. Zeeman spin-splitting is pronounced for $B_{\parallel}$ and negligible for $B_{\perp}$. (c-d) QPC $[001]$ with magnetic field parallel to the direction of current $B_{\parallel}$ and perpendicular $B_{\perp}$ respectively. Zeeman spin-splitting is pronounced for $B_{\perp}$ and small for $B_{\parallel}$. (e-f) QPC $[010]$ with magnetic field parallel to the direction of current $B_{\parallel}$ and perpendicular $B_{\perp}$ respectively. Zeeman spin-splitting is pronounced for $B_{\perp}$ and small for $B_{\parallel}$.}
    \label{fig2}
\end{figure}
  
\twocolumngrid


\textit{Results}
The SOI can be studied by measurements of the $g$-factor, which is extracted from Zeeman spin-splitting measurements. Zeeman spin splitting is studied for in-plane magnetic fields applied either parallel or perpendicular to the current flow in the QPC, and then extracting $g_{\parallel}$ and $g_{\perp}$ from the transconductance $dG/dV_g$. Greyscale maps of the transconductance, showing the 1D subbands evolving in magnetic field, are presented for each QPC in Figure {\color{blue}2}. Dark regions correspond to conductance plateaus and light regions correspond to risers which mark the 1D subband edges. For each QPC the size of the Zeeman spin splitting is dependent on the orientation of the magnetic field with respect to QPC direction. To characterise the SOI we measure the in-plane $g$-factor anisotropy via the ratio $g_\perp/g_\parallel$. This ratio is independent of any multiplicative renormalizations of the $g$-factors such as the voltage to energy conversion individual to each device (lever arm) and Coulomb interaction effects in the QPC channel ~\cite{VionnetPRL16}. This makes the ratio $g_\perp/g_\parallel$ a reliable probe of the SOI that can easily be compared with theory.

The $g$-factors are extracted for the second subbands and higher as the apparent Zeeman spin-splitting is suppressed due to the interplay of SOI and interaction effects in the first subband~\cite{HudsonNatComm21}. For  QPC [011] linear Zeeman spin-splitting of the 1D subbands is observed with the magnetic field $\bm{B}\parallel\bm{I}$ ($g_{\parallel} > 0$) in panel Fig.{\color{blue}~2a}. In contrast nearly no Zeeman spin splitting is observed for $\bm{B}\perp\bm{I}$ ($g_{\perp} \approx 0$) in Fig.{\color{blue}~2b}. Our observation that $g_{\parallel}>g_{\perp}$ is consistent with previous studies of Zeeman spin-splitting in hole QPCs oriented along the $[011]$ crystal axis~\cite{KlochanNJP09, ChenNJP10, KomijaniEPL13, SrinivsasanPRL17}, and is a direct consequence of the strong SOI in hole systems~\cite{MiserevPRL17}. 
    
Surprisingly, we observe the reverse situation in $[001]$ and $[010]$ QPCs oriented $\SI{45}{\degree}$ away from the $[011]$ QPC in either direction. For both $[001]$ QPC, see Fig.{\color{blue}~2c,d}, and $[010]$ QPC, see Fig.{\color{blue}~2e,f}, the Zeeman splitting in $\bm B \parallel \bm I$ is substantially smaller than in $\bm B \perp \bm I$ with the in-plane $g$-factor ratio $g_\perp/g_\parallel \approx 2$. Strong anisotropy of the ratio $g_\perp/g_\parallel$ ranging over one order of magnitude from $g_\perp/g_\parallel \approx 0.15$ for the $[011]$ QPC to $g_\perp/g_\parallel \approx 2$ for the $[001]$ and $[010]$ QPCs cannot be understood within the previous theoretical models where the crystal lattice anisotropy effects were omitted ~\cite{MiserevPRL17, SrinivsasanPRL17, ZulickePSS06, NichelePRL14}. 

\textit{Theory}
For bulk semiconductors, hole dynamics is usually described using the Luttinger Hamiltonian in the spherical approximation~\cite{LuttingerPR56}. The Hamiltonian for holes takes the form
    \begin{equation}
        H_L = \left(\gamma_1 + \frac{5}{2}\bar{\gamma}_2\right) \frac{\mathbf{p}^2}{2m} - \frac{\bar{\gamma}_2}{m}\left(\mathbf{p}\cdot\bm{S}\right)^2
    \end{equation}
where $\mathbf{p}$ is the 3D hole momentum, $\bm{S}$ is the spin $S = 3/2$, $\gamma_1,\bar{\gamma}_2 = \left(2\gamma_2 + 3\gamma_3\right)/5$ are Luttinger parameters, $m$ is the free electron mass~\cite{BaldereschiPRB73}. For a system where holes are confined to a 2D plane, it is convenient to assume radial symmetry with an axis along the direction of confinement, and therefore use an axial approximation of the Luttinger Hamiltonian. In our QPCs, valence band holes are confined along the $z$-axis by the single heterojunction potential (width $\approx \SI{15}{\nano\meter}$), and in the lateral $y$-direction by split QPC gates of width $\SI{300}{\nano\meter}$. 1D ballistic conductance is along the unconfined $x$-direction. The separation of the in-plane $x$-,$y$- and out-of-plane $z$-direction scales allows us to treat a QPC as a planar system that can be described with the effective Hamiltonian approach developed in Ref. \cite{MiserevPRB17}. 
Quantum confinement of holes in 2D quantum wells results in the heavy hole-light hole splitting separating the heavy holes with spin projections $S_z = \pm 3/2$ along the $z$ axis orthogonal to the 2DHG plane and light holes with $S_z = \pm 1/2$.
The lowest 2D subband of 2DHG is comprised of the heavy holes and can be described by an effective 2D Hamiltonian that only depends on the in-plane momentum $\bm k = (k_x, k_y)$~\cite{MiserevPRB17}.
Two heavy hole spin states with $S_z = \pm 3/2$ form the up-down basis for the effective heavy hole pseudospin $1/2$ $\bm{\Sigma} = \boldsymbol{\sigma}/2$, $\boldsymbol{\sigma}$ are the Pauli matrices.
The most important difference of the heavy hole pseudospin $1/2$ from the electron spin $1/2$ is that a flip of the heavy hole pseudospin requires three quanta of angular momentum (to promote $S_z = - 3/2$ to $S_z = +3/2$) instead of a single quantum for electron spin.
Thus, the raising and lowering heavy hole spin operators $\Sigma_\pm = \Sigma_x \pm i \Sigma_y$ carry three quanta of angular momentum.
Other vector operators, e.g. $k_\pm = k_x \pm i k_y$ and $B_\pm = B_x \pm i B_y$, carry a single quantum of angular momentum.
These angular momentum selection rules allow for two different spin-orbital terms describing the in-plane Zeeman splitting in 2DHG in absence of the crystal lattice anisotropy~\cite{MiserevPRB17}:
    \begin{equation} \label{MiserevHamiltonian}
        H_0 = -\frac{g_1(k)}{2 k^2} \Sigma_+ B_- k^2_-  -\frac{g_2(k)}{2 k^4} \Sigma_+ B_+ k^4_- + h.c. 
    \end{equation}
where $k = |\bm k|$, $k_\pm = k_x \pm i k_y$, $B_{\pm} = B_x \pm i B_y$, $\Sigma_{\pm} = \Sigma_x \pm i \Sigma_y$, $h.c.$ stands for hermitian conjugate.
The Hamiltonian $H_0$ valid at arbitrary $\mathbf{k}$ indicated by the $k$-dependence of the dimensionless $g$-factors $g_{1,2}(k)$ that have to be calculated numerically for a given heterostructure confinement. 
The $1/k^2$ and $1/k^4$ normalisation of the $g_{1,2}(k)$ in Eq.~\eqref{MiserevHamiltonian} is chosen for convenience. The effective Hamiltonian $H_0$ in Eq.~\eqref{MiserevHamiltonian} explains the different values for $g_\perp$ and $g_\parallel$ observed in $[011]$ and $[01\bar{1}]$ QPCs (references) yet it is not enough to explain the strong dependence of the ratio $g_\perp/g_\parallel$ on the QPC orientation, see Fig.{\color{blue}~2,~4b}.

We now step beyond the axial approximation and take into account the cubic symmetry of the lattice (for now neglecting the more subtle contribution of bulk inversion asymmetry, known as Dresselhaus SOI~\cite{DresselhausPR55}). Zinc-blende $(100)$ heterostructures have square symmetry i.e. the effective Hamiltonian is invariant under $\pi/2$ rotations. The square symmetry only allows terms in the effective Hamiltonian that carry multiples of the angular momenta $\pm 4$ e.g. $j_z = 0,\pm 4,\pm 8..$. However high angular momentum harmonics $j_z = \pm 4N$ are suppressed by the factor $\eta ^N$ where $\eta = (\gamma_3 - \gamma_2)/(\gamma_3 + \gamma_2)$, and $\gamma_{2,3}$ are Luttinger parameters. In GaAs $\eta \approx 0.17$ which allows us to take into account only the $N = 0,1$ harmonics. The following are all the possible terms in the effective Zeeman Hamiltonian linear with respect to the in-plane magnetic field $\bm{B}$ and the heavy hole pseudospin $\bm{\sigma}$ that carry angular momenta $\pm 4$:
    \begin{equation} \label{Hamiltonian}
    \begin{split}
        H_1 = &-\frac{g_3(k)}{2}\left(S_+ B_+ + h.c.\right)
         - \frac{g_4(k)}{2k^2}\left(S_+ B_- k^2_+ + h.c.\right) \\
        & - \frac{g_5(k)}{2k^6}\left(S_+ B_- k^6_- + h.c.\right) 
         - \frac{g_6(k)}{2k^8}\left(S_+ B_+ k^8_- + h.c.\right) 
    \end{split}
    \end{equation}
where $g_i(k)$ are dimensionless couplings that depend only on $k = |\bm k|$. Terms $g_{3,4}$ have been previously calculated in the limit of small $\bm{k}$~\cite{KomijaniEPL13}. Here we do not imply that $\bm{k}$ is small. Eq.~\eqref{Hamiltonian} is written in cubic crystal coordinates with $X = [010]$, $Y = [001]$. The terms in $H_1$ will pick up additional phases $e^{\pm 4 i \phi}$ in a coordinate system rotated by angle $\phi$ (see Supplementary Material).
    
For numerical calculations of $g_i(k)$, we use a triangular potential profile (approximating a single GaAs/AlGaAs heterojunction as in our experimental device) self-consistently screened by the finite hole density $p = 2\times 10^{11}\SI{}{\per\square\centi\meter}$, see Fig.{\color{blue}~S4} in the Supplementary Material. 
At the Fermi momentum $k_F = (2\pi p)^{1/2} = 1.12\times 10^{-2}\SI{}{\per\angstrom}$ corresponding to the 2D hole density $p = 2\times 10^{11}\SI{}{\per\square\centi\meter}$ 
the $g$-factors $g_i(k)$ cannot be reliably approximated by the small-$k$ expansion.
In our experimental case of a more asymmetric quantum well with strong Rashba SOI, $g_1$ is the dominant contribution. $g_2 \approx -0.05$ is strongly suppressed by the Rashba SOI. However, $g_2 \sim g_1$ in square quantum wells where the Rashba SOI is much smaller~\cite{MiserevPRL17}. 
$g_3 \approx - 0.25$ and $g_4 \approx -0.18$ are responsible for the dependence of the overall $g$-factor on QPC orientation with respect to the crystal axes. $g_5 \approx -0.05$ and $g_6 \approx -0.03$ are still small but they become significant, $|g_{5,6}|>0.1$, at high densities $p > 3\times 10^{11}\SI{}{\per\square\centi\meter}$ which is beyond what can be presently achieved in our QPC devices (see Supplementary Material).

From eqs.~\eqref{MiserevHamiltonian} and~\eqref{Hamiltonian}, we write an expression for the overall in-plane $g$-factor as the sum of each contributing $g_i$ term. The $g_{1}$ and $g_{2}$ terms derived in the axial approximation depend on the magnetic field orientation $\theta$ with respect to the QPC. The $g_{3}$, $g_{4}$, $g_{5}$ and $g_{6}$ terms arise from the angular momenta $\pm 4$ permitted by the square symmetry of the $(100)$ zinc-blende crystal, and therefore depend on the magnetic field orientation with respect to the QPC $\theta$ \emph{and} the QPC orientation with respect to the crystal axes $\phi$. 
According to Ref.~\cite{MiserevPRL17}, $\langle n| k_\perp^2 |n \rangle \approx k_F^2$ and $\langle n | k_\parallel^2 |n \rangle \ll k_F^2$ at the edge $|n \rangle$ of a high $n \ge 2$ QPC subband due to strong electrostatic flattening of a QPC channel. Here $k_\perp$ and $k_\parallel$ are the hole momentum operators perpendicular and parallel to the QPC channel,  allowing for the quasiclassical treatment of the hole momentum at the edge of $n^{\rm th}$, $n \ge 2$, QPC subband i.e. $\langle n | g_i(k)|n \rangle \approx g_i(k_F)$.
Thus, we can extract the in-plane hole $g$-factor of the $n \ge 2$ QPC subbands by applying the quasiclassical approximation to the 2D Zeeman Hamiltonian $H = H_0 + H_1$:
    \begin{multline} \label{gfactor}
        |g\left(\theta,\phi\right)| = |-g_1 e^{-i\theta} + g_2 e^{i\theta} + g_3 e^{i(\theta + 4\phi)} \\ - g_4 e^{-i(\theta - 4\phi)} - g_5 e^{-i(\theta + 4\phi)} + g_6 e^{i(\theta - 4\phi)}|
    \end{multline}
where $g_i = g_i(k_F)$. In the coordinate system which is rotated by angle $\phi$ with respect to the crystal axes $X = [010]$, $Y = [001]$, the $k^2$ operator has only $k_y^2$ average at the subband edge where $k_x^2$ is vanishing, hence $k_+^2 = k_-^2 = -k_y^2 \approx -k_F^2$.
    
Equation~\eqref{gfactor} allows us to plot the $g$-factor parallel to the QPC current $g_{\parallel}$, and in-plane perpendicular $g_{\perp}$, as a function of QPC orientation with respect to crystal axes $\phi$. The $g$-factor parallel to current $g_{\parallel}$ corresponds to $\theta = 0$, $g_{\parallel} (\phi) = |g(0,\phi)|$, and perpendicular to current $g_{\perp}$ corresponds to $\theta = \pi/2$, $g_{\perp} (\phi) = |g(\pi/2, \phi)|$. $|g_{\parallel}(\phi)|$ and $|g_{\perp}(\phi)|$ are plotted in Figure {\color{blue}3} \color{black} for $p = 2\times 10^{11}\SI{}{\per\square\centi\meter}$. A very strong angular dependence of $g_\perp$ is observed, varying from $g_\perp \approx 0.1$ for $[011]$ and $[01\bar{1}]$ QPCs to $g_\perp \approx 1.2$ for $[001]$ and $[010]$ QPCs, and nearly isotropic $g_\parallel \approx 0.6$ for any QPC orientation. The anisotropy of the in-plane $g$-factors for the $[011]$ and $[01\bar{1}]$ crystal directions (corresponding to $\phi =\pi/4, 3\pi/4$) is $g_{\parallel} > g_{\perp}$. In contrast, for the $[001]$ and $[010]$ crystal directions (corresponding to $\phi = 0, \pi/2$) $g_{\perp}$ becomes large, while $g_{\parallel}$ is only slightly suppressed and the anistropy of the in-plane $g$-factors is reversed $g_{\parallel} < g_{\perp}$. 

Experimental $g$-factors extracted from Fig.{\color{blue}~2} for all three QPCs for different subbands $n \ge 2$ show nearly no subband dependence (within the error bars), see Fig.{\color{blue}~S1} in the Supplementary Material. This is consistent with electrostatic calculations performed in Ref.~\cite{MiserevPRL17} where the QPC $g$-factors saturate to (nearly) same value already at $n \ge 2$. Therefore, we average the $g$-factors over the subband index $n \ge 2$ to increase experimental accuracy. Averaged $g$-factors for each QPC are shown in Figure {\color{blue}4a}. We see that there is a substantial difference between the $[010]$ and $[001]$ QPCs which likely comes from the multiplicative renormalizations of the $g$-factor originating from the voltage-to-energy conversion and the details of the Coulomb interaction~\cite{MiserevPRL17, VionnetPRL16}. These renormalizations are cancelled out if we consider the ratio $g_\perp/g_\parallel$ shown in Fig.{\color{blue}~4b} where we find that this ratio is indeed the same for $[010]$ and $[001]$ QPCs as expected. The dashed line in Fig.{\color{blue}~4b} shows the theoretical dependence of $g_\perp(\phi)/g_\parallel(\phi)$, where $g_\perp(\phi)$ and $g_\parallel(\phi)$ are plotted in Fig.{\color{blue}~3}. The theoretical prediction for $g_\perp/g_\parallel$ is thus fully consistent with observed ratios $g_\perp/g_\parallel \approx 0.15$ in $[011]$ QPC and $g_\perp/g_\parallel \approx 2$ in $[010]$ and $[001]$ QPCs, see Fig.{\color{blue}~4b}

\begin{figure}[ht]
    \centering
    \includegraphics[width=0.45\textwidth]{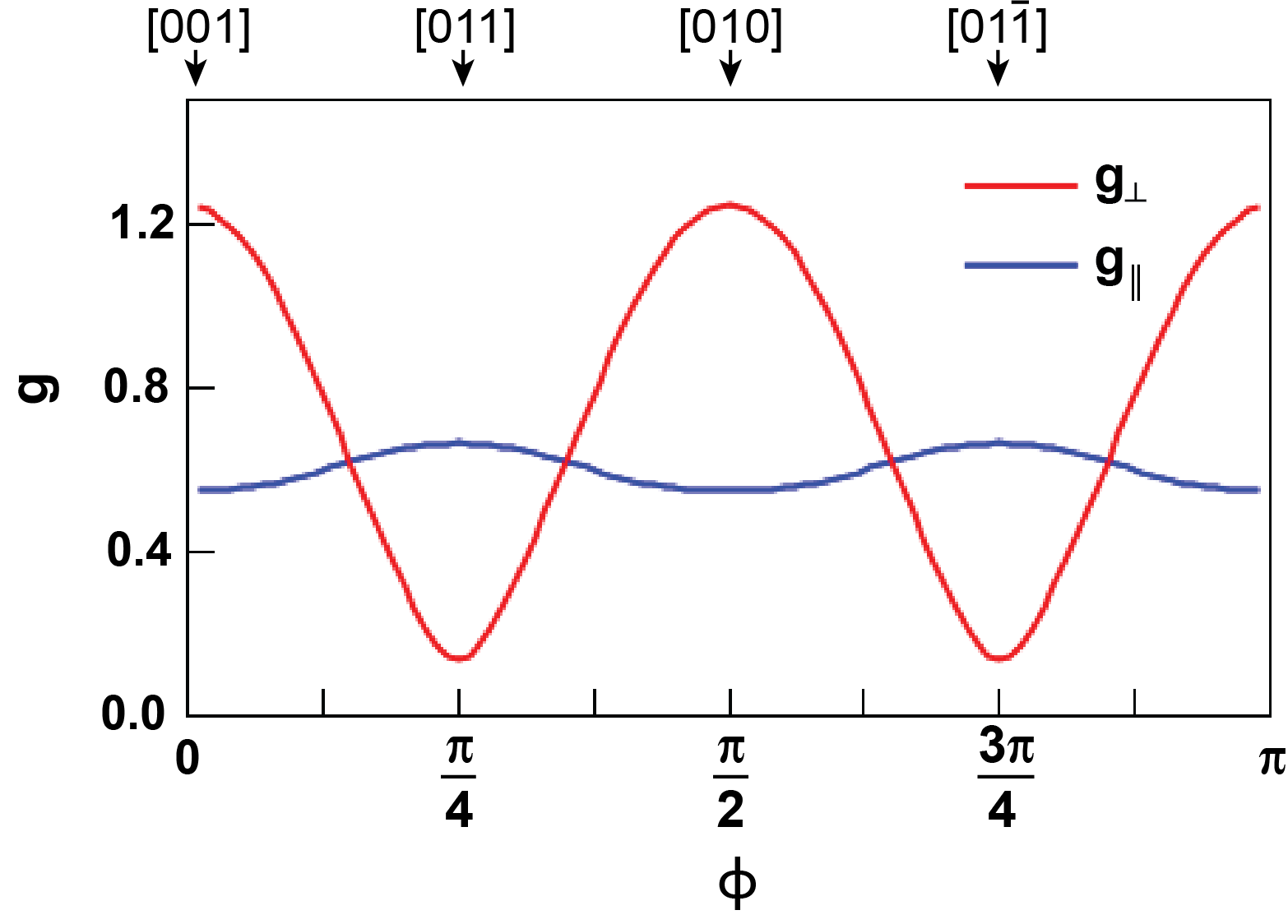}
    \caption{Numerically calculated $g_{\parallel}$- and $g_{\perp}$-factors as a function of QPC orientation with respect to crystal axes ($\phi$). $\phi = 0, \pi /2$ correspond to crystal directions $[001]$ and $[010]$ respectively. $\phi = \pi /4$ corresponds to crystal direction $[011]$.}
    \label{fig3}
\end{figure}

\begin{figure}[ht]
    \centering
    \includegraphics[width=0.45\textwidth]{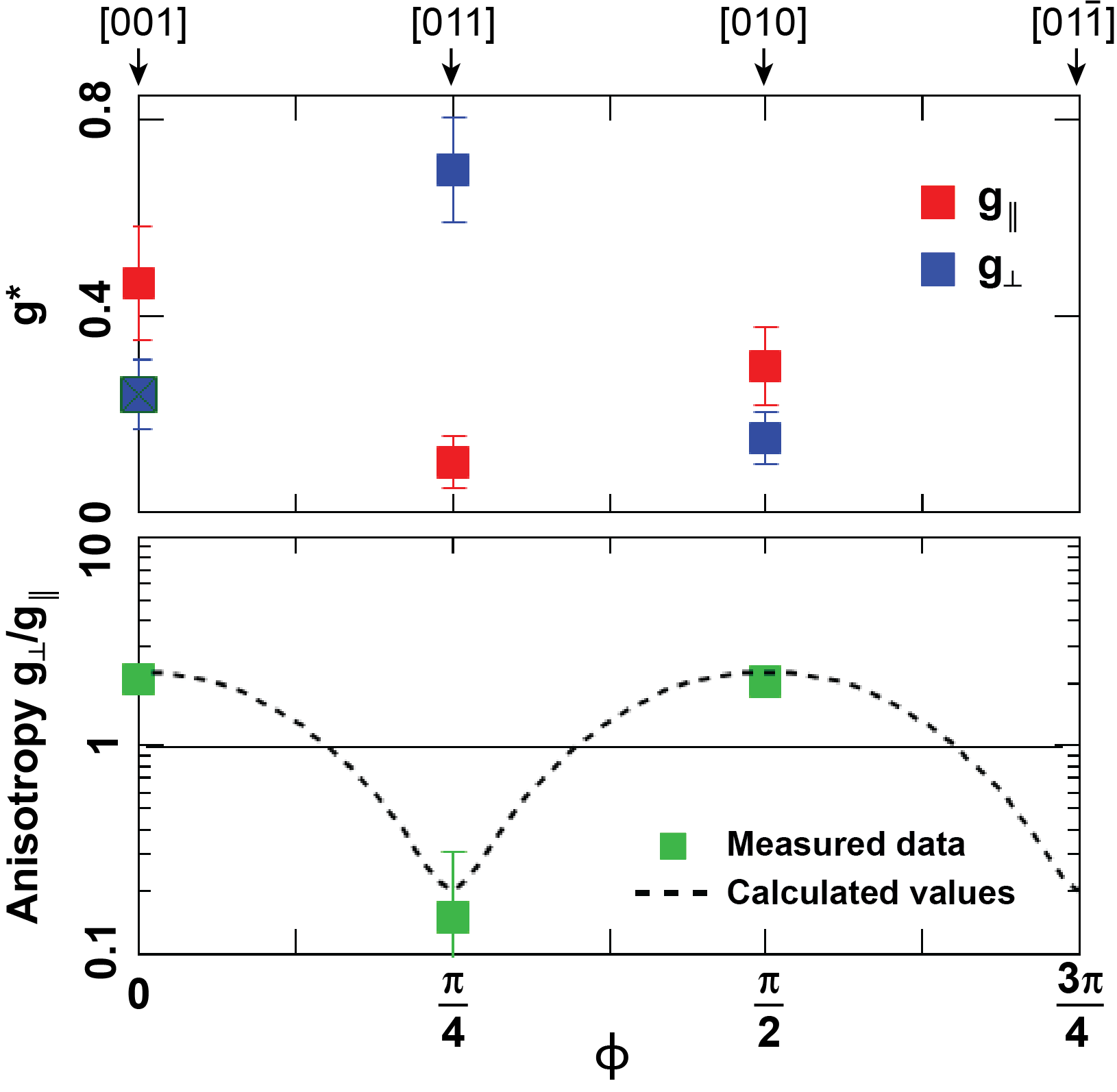}
    \caption{(a) Experimentally obtained $g$-factors calculated from the Zeeman spin-splitting measurements in Fig.{\color{blue}~2}\color{black}. The $g$-factor for each 1D subband is averaged to obtain a single $g_{\parallel}$- and $g_{\perp}$-factor for each value of $\phi$. (b) $g$-factor anisotropy $g_\perp /g_\parallel$ plotted as a function of QPC orientation with respect to crystal axes $\phi$. Dashed line is numerically calculated anisotropy of the in-plane $g$-factors. Green squares are experimental data points calculated from the data in panel (a). $\phi = 0, \pi /2$ correspond to crystal directions $[001]$ and $[010]$ respectively. $\phi = \pi /4$ corresponds to crystal direction $[011]$.}
    \label{fig4}
\end{figure}

\textit{Conclusion}
We observed for the first time very strong crystal lattice anisotropy of the in-plane Zeeman splitting in $(100)$ GaAs hole QPCs. The ratio $g_\perp/g_\parallel$ of the in-plane $g$-factors measured perpendicular, $g_\perp$,
and parallel, $g_\parallel$, to the QPC channel varies by an order of magnitude from $g_\perp/g_\parallel \approx 0.15$ in $[011]$ QPC to $g_\perp/g_\parallel \approx 2$ in $[010]$ and $[001]$ QPCs.
The dramatic dependence of $g_\perp/g_\parallel$ on the QPC orientation signals strong crystal lattice anisotropy of the in-plane Zeeman splitting.
To understand this effect, we derived the effective 2D Hamiltonian describing the in-plane Zeeman splitting in 2DHG.
Additionally to the axially symmetric terms $g_1$ and $g_2$ ($k^2$ and $k^4$ in Eq.~\eqref{MiserevHamiltonian}), there is the crystal lattice correction, $g_3$, $g_4$, $g_5$, and $g_6$ ($k^0$, $k^2$, $k^6$, $k^8$ in Eq.~\eqref{Hamiltonian}), which results in the angular dependence of the hole $g$-factors on the QPC orientation shown in Fig.{\color{blue}~3}. Performing numerical calculations for realistic parameters of the wafer, we find excellent agreement between theoretical and experimental values of $g_\perp/g_\parallel$ shown in  Fig.{\color{blue}~4b}. Our work shows that the crystal lattice anisotropy is very strong in GaAs 2D hole systems which provides additional tunability for hole-based spin devices.


\begin{acknowledgements}
This work was supported by the Australian Research Council under the Discovery Projects scheme, and was performed in part using facilities of the NSW Node of the Australian National Fabrication Facility. D.A.R and I.F. acknowledge support from the Engineering and Physical Sciences Research Council, United Kingdom. K.L.H. acknowledges support from Sydney Quantum Academy. D.S.M acknowledges the support by the Georg H. Endress Foundation and NCCR SPIN.
\end{acknowledgements}


%

\end{document}